\documentclass[amsmath,amssymb,prb,floatfix,twocolumn,superscriptaddress]{revtex4-1}

\usepackage{graphicx}
\usepackage{dcolumn}
\usepackage{bm}
\usepackage{times}
\usepackage{colordvi,epsf}
\usepackage{color}
\usepackage[normalem]{ulem}
\newcommand{\vcn}[1]{\hat{#1}}                        
\newcommand{\vcc}[1]{{\boldsymbol{#1}}}               
\newcommand{\matcc}[1]{\underline{\underline{\mathcal{#1}}}} 
\newcommand{\abs}[1]{\lvert#1\rvert}                  

\renewcommand{\eqref}[1]{(\ref{#1})}

\newcommand{\eg}{e.g.\@ }
\newcommand{\ie}{i.e.\@ }
\newcommand{\cf}{cf.\@ }
\newcommand{\etal}{\textit{et al.\@ }}

\newcommand{\mbohr}{\mu_\mathrm{B}} 

\begin{document}

\title{Comparison of first-principles methods to extract magnetic parameters in ultra-thin films: Co/Pt(111)}

\author{Bernd Zimmermann}
\email{be.zimmermann@fz-juelich.de}
\affiliation{Peter Gr\"unberg Institut and Institute for Advanced Simulation, Forschungszentrum J\"ulich and JARA, 52425 J\"ulich, Germany}

\author{Gustav Bihlmayer}
\affiliation{Peter Gr\"unberg Institut and Institute for Advanced Simulation, Forschungszentrum J\"ulich and JARA, 52425 J\"ulich, Germany}

\author{Marie B\"ottcher}
\affiliation{Institute of Physics, Johannes Gutenberg-Universit\"at, 55128 Mainz, Germany}
\affiliation{Graduate School Materials Science in Mainz, 55128 Mainz, Germany}

\author{Mohammed Bouhassoune}
\affiliation{Peter Gr\"unberg Institut and Institute for Advanced Simulation, Forschungszentrum J\"ulich and JARA, 52425 J\"ulich, Germany}

\author{Samir Lounis}
\affiliation{Peter Gr\"unberg Institut and Institute for Advanced Simulation, Forschungszentrum J\"ulich and JARA, 52425 J\"ulich, Germany}

\author{Jairo Sinova}
\affiliation{Institute of Physics, Johannes Gutenberg-Universit\"at, 55128 Mainz, Germany}
\author{Stefan Heinze}
\affiliation{Institute of Theoretical Physics and Astrophysics, Christian-Albrechts-Universit\"at, 24098 Kiel, Germany}

\author{Stefan Bl\"ugel}
\affiliation{Peter Gr\"unberg Institut and Institute for Advanced Simulation, Forschungszentrum J\"ulich and JARA, 52425 J\"ulich, Germany}

\author{Bertrand Dup\'e}
\affiliation{Institute of Physics, Johannes Gutenberg-Universit\"at, 55128 Mainz, Germany}

\date{\today}

\begin{abstract}
We compare three distinct computational approaches based on first-principles calculations within density functional theory to explore the magnetic exchange and the Dzyaloshinskii-Moriya interactions (DMI) of a Co monolayer on Pt(111), namely (i) the method of infinitesimal rotations of magnetic moments based on the Korringa-Kohn-Rostoker (KKR) Green function method, (ii) the generalized Bloch theorem applied to spiraling magnetic structures and (iii) supercell calculations with non-collinear magnetic moments, the latter two being based on the full-potential linearized augmented plane wave (FLAPW) method. In particular, we show that the magnetic interaction parameters entering micromagnetic models describing the long-wavelength deviations from the ferromagnetic state might be different from those calculated for fast rotating magnetic structures, as they are obtained by using (necessarily rather small) supercell or large spin-spiral wave-vectors.
In the micromagnetic limit, which we motivate to use by an analysis of the Fourier components of the domain-wall profile, we obtain consistent results for the spin stiffness and DMI spiralization using methods (i) and (ii). The calculated spin stiffness and Curie temperature determined by subsequent Monte Carlo simulations are considerably higher than estimated from the bulk properties of Co, a consequence of a significantly increased nearest-neighbor exchange interaction in the Co-monolayer (+50\%). The calculated results are carefully compared with the literature.
\end{abstract}

      
\maketitle

\section{Introduction}

In recent years, non-collinear magnetic structures and in particular skyrmions, have attracted a lot of interest due to their peculiar properties and their technological perspective in the field of information technology~\cite{nnano.2013.29}. Typically these magnetic structures are stabilized by the competition between the Heisenberg exchange, magnetic anisotropy and dipolar interaction. Recently, the Dzyaloshinskii-Moriya interaction (DMI) \cite{Dzialoshinskii1957,PhysRev.120.91} has emerged as a new key stabilization mechanism. The DMI arises due to spin-orbit coupling (SOC) and is present in every system, which lacks structural inversion symmetry. The presence of DMI explains the stabilization of skyrmions in bulk B20 alloys such as MnSi~\cite{muhlbauer2009skyrmion} or in thin films of semiconductor Fe$_{(1-x)}$Co$_x$Si~\cite{Munzer2010} and in ultra-thin films at low temperature such as in Fe/Ir(111)~\cite{nphys2045} or Pd/Fe/Ir(111)~\cite{Romming2013}.

As skyrmions are becoming relevant for technological applications~\cite{Zhang2015a}, additional design goals for skyrmions have been formulated \cite{Fert2017}. For example, (i) they should be stable above room temperature, (ii) skyrmions should not be too small ($\gtrsim 5$~nm diameter), (iii) skyrmions in ultra-thin films and heterostructures thereof are preferred over skyrmions in bulk samples, and (iv) preferably use materials that are simple to integrate into current manufacturing processes. The latter brings Co/Pt(111) ultra-thin film into play, a material that is well known and is used for perpendicular magnetic recording~\cite{Yang2015a,Moreau-Luchaire2016,Boulle2016}. Several recent studies focus on Co/Pt based systems\cite{Maccariello2018,Chauleau2018} and explore the possibility to tune the material parameters, \eg through additional buffer layers \cite{Vida2016,Simon2018,Jia2018}, alloying \cite{Shahbazi2018,Hanke2018} or dusting \cite{Zimmermann2018} with a third chemical element.

Obviously, it becomes crucial to understand the stabilization mechanism of skyrmions, predict and design their properties in setups for technological use by theoretical models. {\it Ab initio} spin-lattice models proved to be a very powerful approach to realistically describe non-collinear magnets, single skyrmions and skyrmions lattices in experimentally realized systems~\cite{Dupe2014,Nandy2016,Dupe2016b}. In such a model, magnetic moments are localized at atomic sites and their interactions are described by parameters, typically the Heisenberg exchange constants, $J_{ij}$, and the DMI vector, $\mathbf{D}_{ij}$, for two-site interactions ($i$ and $j$ label magnetic sites), or the magnetic on-site anisotropy $K_i$. The parameters for an {\it ab initio} spin-lattice model are obtained directly from the total energy of the electronic structure by density functional theory, and the magnetic ground state is found for example by spin-dynamics or Monte-Carlo methods.

While this multiscale approach provides a very efficient description of the energy landscape, the quality of the description depends crucially on the parameters obtained by the mapping of the density functional description of the magnetic states onto the model. In this respect it is important to notice that almost all bulk and interface stabilized skyrmion systems are itinerant magnets, \ie those electrons responsible for the formation of magnetism also participate in the formation of a complex Fermi surface and hop across the lattice. As a consequence the magnetic interactions are typically not short ranged, as it is often assumed in spin-lattice models. Additionally, the size of the magnetic moments, $M$, is not an integer multiple of the Bohr magneton, but depends on the details of the electronic structure. Most importantly, and in difference to the basic assumptions of many spin models, the size of the moments and the interaction parameters between them depend on their relative orientation. This effect increases with the number of magnetic neighbors and is thus stronger for itinerant bulk magnets than for ultra-thin films. The effect of the change of the magnetic moment for a spin-spiral state with wave vector $\vcc{q}$ imposed onto the bulk magnets Cr, Mn and Fe was shown \cite{Bihlmayer_DFT_of_magnetism}. In the case of Ni, the magnetic moment can even be completely quenched in the antiferromagnetic state which leads to an overestimation of the total energy~\cite{Singer2005}. Similarly, the magnitude of induced magnetic moments strongly depends on the spin-configuration of neighboring strong moments~\cite{Polesya2010}. As a consequence the range of validity of the {\it ab initio} spin-lattice model depends crucially on the initial spin configuration for which the parameters entering the spin models have been calculated. The choice of the spin-configuration depends then on the purpose of the application of the spin-lattice model for example to explore the ground states at low or high temperatures or the excited states of small or large spin structures.

In case of the skyrmions, and in particular for skyrmions in the Co thin films on 5\emph{d} substrates~\cite{Boulle2016,Corredor2017,Perini2018} or Co/Ru(0001)~\cite{Herve2017}, one deals with relatively large non-collinear magnetic textures, with sizes in the order of 100~nm. In this case, the variation of the angle between magnetic moments is small from one atom to the next and the ferromagnetic state (FM) is a good initial state to determine the model parameters. Nevertheless, it is  interesting to know how sensitive the model parameters are with respect to the initial state.

In this work, we compare three different approaches to extract the model parameters, namely (i) by the method of infinitesimal rotations based on the Korringa-Kohn-Rostoker (KKR) Green function method, or by mapping the energies of spin-spiral states which are calculated (ii) by using the generalized Bloch theorem (gBT) or (iii) in a supercell geometry. The latter two approaches employ the full potential linearized augmented plane wave (FLAPW) method. We obtain consistent results for the micromagnetic spin stiffness and DMI spiralization in the long-wavelength limit (\ie around the ferromagnetic state) between KKR and gBT calculations. For larger spin-spiral vectors, the details of the computational procedure (\eg whether flat or coned spin-spirals are used) become relevant. As a consequence, relying only on one data-point is to be taken with caution, as it is typically done in supercell calculations. Our calculated spin-stiffness is considerably higher (more than 50\%) as compared to experimentally determined values for Co-thicknesses below 1~nm. We also find a considerably higher Curie temperature as compared to an estimate based on the Curie temperature of bulk Co., which we trace back to an increased nearest-neighbor exchange interaction in the monolayer as compared to bulk Co.

The paper is structured as follows: in Section~\ref{sec:methods} we describe the structure of the Co/Pt(111) system that we study, as well as the magnetic models for which we extract the parameters and detail the computational approaches used. In Section~\ref{sec:results} we present the results and discuss them in comparison to the existing literature, before we conclude in Section~\ref{sec:conclusions}. The Appendices investigate the dependence of the predicted Curie temperature on the computational procedure (Appendix~\ref{appendix:MonteCarlo}), give arguments why the micromagnetic limit is the suitable one for Co/Pt based systems (Appendix~\ref{appendix:lengthscales}) and investigate the importance of the induced Pt moments for the spin-spiral energy dispersion (Appendix~\ref{appendix:inducedmoments}).

\section{Methods and Computational Details}
\label{sec:methods}

We have studied a Co monolayer on Pt(111) in both fcc and hcp stacking positions by means of density functional theory (DFT) calculations. For all calculations, the calculated in-plane lattice parameter of bulk fcc-Pt in local-density approximation (LDA) \cite{doi-10.1139/p80-159} was used ($a_\text{fcc}=0.390$~{nm}). We use a two-dimensional setup, \ie embedding a finite number of layers between two semi-infinite vacuum regions.

\subsection{Structural Relaxations}
\label{section:structuralRelaxations}

The structural relaxations were performed using the DFT package FLEUR~\cite{FLEUR} employing the full potential linearized augmented plane wave (FLAPW) method. We used a mixed density functional, which was introduced in Ref.~\onlinecite{mixed-fun} to treat combined systems of 3$d$- and 5$d$-transition metals. It combines the LDA in the muffin tin (MT) spheres of the 5$d$ atoms and the generalized gradient approximation (GGA) \cite{PhysRevB.46.6671} everywhere else. It was already used to obtain accurate description of surfaces and interfaces containing 3$d$ and 5$d$ materials~\cite{Dupe2014,Dupe2015}. We used a cutoff parameter for the basis functions of $K_\mathrm{max}=4.0~a_\mathrm{B}^{-1}$ and 72 $k$-points in one twelfth of the two-dimensional Brillouin zone (BZ), where $a_\mathrm{B}$ is the Bohr radius. The symmetric slab was composed of five Pt layers and one Co layer on each side. The positions of Co and the top Pt layers were relaxed for both Co stacking positions. As shown in Table~\ref{tab:relaxation}, the relaxed structural parameters are basically independent of the Co stacking. Our calculations suggest that the ground state of a Co monolayer on Pt(111) is obtained for the fcc stacking position. Hence, in the analysis presented in Sec.~\ref{sec:results}, we focus mostly on the fcc stacking position.

\begin{table}[thp]
\centering
\caption{The used in-plane lattice parameter $a$ and relaxed interlayer distances $d$ for two different stacking positions of the Co monolayer on Pt(111). Distances are given in \AA\ and total energies in meV/(Co atom) relative to the fcc stacking position.}
\label{tab:relaxation}
\begin{ruledtabular}
\begin{tabular*}{\hsize}{c@{\extracolsep{0ptplus1fil}}cc}
\multicolumn{3}{c}{Co/Pt(111)}  \\[0.4ex] \hline \hline
{}                          &   fcc  &   hcp  \\[0.4ex] \hline
$a$          			 	&  2.76  &  2.76  \\
$d_{\text{Co-Pt1}}$         &  2.02  &  2.03  \\
$d_{\text{Pt1-Pt2}}$        &  2.37  &  2.38  \\
Total energy                &  0     &  134 
\end{tabular*}
\end{ruledtabular}
\end{table}

\subsection{Magnetic models}

\subsubsection{Extended Heisenberg model}

Magnetic ultrathin films are well described by the general atomistic extended Heisenberg Hamiltonian,
\begin{eqnarray}
	H &=& -\sum_{i,j} J_{ij} ~ \vcc{m}_i \cdot \vcc{m}_j + \sum_{i,j} \vcc{D}_{ij} \cdot \left( \vcc{m}_i \times \vcc{m}_j \right) \label{eq:model_spinlattice} \\ \nonumber
	{} & + & \sum_{i} K_{i} \left(m_i^z\right)^2
\label{Eq:H}
\end{eqnarray}
where $\vcc{m}_i$ and $\vcc{m}_j$ are the magnetic moments of unit length at position $\vcc{R}_i$ and $\vcc{R}_j$ respectively,  $J_{ij}$ are the magnetic exchange parameters,  $\vcc{D}_{ij}$ are the Dzyaloshinskii-Moriya vectors and $K_{i}$ is the onsite uniaxial magnetocrystalline anisotropy. 

The extended Heisenberg Hamiltonian associates magnetic moments to atomic sites. It was \textit{a priori} not designed to study itinerant magnets. However, the inclusion of exchange and DM energy parameters beyond the first nearest neighbor approximation allows for an accurate description of the energy landscape of even frustrated 2-dimensional itinerant magnets on Ir(111)~\cite{nphys2045,Simon2014,Dupe2014} and Ir(001)~\cite{Hoffmann2015} and on W(110) substrates~\cite{Hoffmann2017}. 

A particularly important subset of non-collinear magnetic states are spin spirals,
\begin{equation}
\vcc{m}_{i} = \matcc{R}(\vcn{n}) ~ \left( \begin{array}{c}
   \sin\theta \, \cos(\vcc{q} \cdot \vcc{R}_{i})  \\
   \sin\theta \, \sin(\vcc{q} \cdot \vcc{R}_{i})  \\
   \cos\theta \, \end{array}  \right),
\label{eq:def_spinspiral}
\end{equation}
with $\vcc{q}$ being the spin-spiral vector, $\vcc{R}_{i}$ is the position of site $i$, the rotation matrix $\matcc{R}(\vcn{n})$ brings the local $z$ axis to the rotation axis $\vcn{n}$ of the spin spiral, and $\theta$ is the cone angle. For the special value $\theta=\pi/2$, we obtain flat spin spirals. Within the spin-lattice model, Eq.~\eqref{eq:model_spinlattice}, the Heisenberg and DM energy contributions of flat spin-spirals are given by
\begin{eqnarray}
    E_\mathrm{SS}(\vcc{q}) &=& \sum_{j} J_{0j} \left[ 1 - \cos\left( \vcc{q} \cdot \vcc{R}_{0j} \right) \right] \label{eq:E_ss_from_Jij} \\
    E_\mathrm{DM}(\vcc{q};\vcn{n}) &=& \sum_{j} \left( \vcn{n} \cdot \vcc{D}_{0j} \right) ~ \sin\left( \vcc{q} \cdot \vcc{R}_{0j} \right) ~, \label{eq:E_DM_from_Dij}
\end{eqnarray}
with $\vcc{R}_{0j} = \vcc{R}_j - \vcc{R}_0$.

One may calculate the $J_{ij}$ and $\vcc{D}_{ij}$ parameters in different ways (see Sec.~\ref{sec:methods:DFT}). In any case, Eq.~\eqref{eq:model_spinlattice} is only valid under the condition that the size of all magnetic moments is constant. In DFT, this constraint does not exist and particular care must be taken in the extraction of magnetic exchange interactions if this condition is not respected~\cite{Lezaic2013, Polesya2010}.

\subsubsection{Micromagnetic and effective model}

The micromagnetic model employs a continuous vector field, the magnetization $\vcc{m}(\vcc{r})$. In case of a thin-film geometry, the micromagnetic energy is given by the functional
\begin{eqnarray}
    E[\vcc{m}] &=& \int \mathrm{d}^3 r ~ \Big\lbrace A ~ (\nabla \vcc{m})^2 - K ~ (\vcc{m} \cdot \hat{z})^2 \\
    && + D \left[ \vcc{m} (\nabla \cdot \vcc{m}) - (\vcc{m} \cdot \nabla) \vcc{m}\right] \cdot \hat{z}  \Big\rbrace \nonumber
\end{eqnarray}
with the exchange stiffness (also termed spin stiffness) $A$, uniaxial anisotropy constant $K$ and interfacial DMI constant (or spiralization) $D$, and $\hat{z}$ is the unit vector along the direction perpendicular to the film. The material parameters $A$ and $D$ are related to the parameters of the atomistic model by \cite{Schweflinghaus2016}
\begin{eqnarray}
    A &=& \frac{1}{2 V_\Omega} \sum_{j} J_{0j} ~ \left(R_{0j}^x\right)^2 \label{eq:stiffness:Jij} \\
    D &=& \frac{1}{V_\Omega} \sum_{j} D^y_{0j} ~ R_{0j}^x ~. \label{eq:spiralization:Dij}
\end{eqnarray}
Here, $V_\Omega$ is the volume of the magnetic part of the unit cell, \ie in our case the Co monolayer. It may become difficult to define the thickness of a layer in the presence of relaxations (see \eg the discussion in Ref.~\onlinecite{Simon2018}). Here, we take $t_\mathrm{Co} = 0.2~\mathrm{nm}$ as thickness of the monolayer (see Table~\ref{tab:relaxation}).

Finally, we can express the micromagnetic parameters as effective parameters of a nearest-neighbor Heisenberg model (see Eq.~\eqref{eq:model_spinlattice}),
\begin{eqnarray}
    J_\mathrm{eff} = \frac{2}{3} \frac{V_\Omega}{a^2} A \\
    D_\mathrm{eff} = \frac{1}{3} \frac{V_\Omega}{a} D~, \label{eq:conversion:Deff+D}
\end{eqnarray}
which is designed to reproduce the energy in the long-wavelength limit, but deviations for other magnetic states are expected.

\subsection{Extraction of magnetic interaction parameters from DFT}
\label{sec:methods:DFT}

\begin{table}[thp]
    \centering
  \caption{Overview of and key differences between methods used in this work for the extraction of magnetic parameters from DFT.}
  \label{tab:methods:comparison}
  \begin{ruledtabular}
  \begin{tabular}{c|cc}
                      & self-consistent             & perturbative  \\[1.1ex] \hline
    KKR               & FM state                    & non-collinearity \\
    FLAPW--gBT        & any spin-spiral $\vcc{q}$   & spin-orbit coupling   \\
    FLAPW--supercell  & large spin-spiral $\vcc{q}$ & ---   \\
    \end{tabular}
    \end{ruledtabular}
\end{table}

In the next step, we describe how to extract the parameters for a spin-lattice model from DFT calculations employing three different approaches, which are briefly described here and more detailed information is given in the rest of this subsection. A summary of main differences is presented in Table~\ref{tab:methods:comparison}
\begin{enumerate}
\item{(KKR) The first approach relies on the KKR method. We perform self-consistent calculation for the ferromagnetic state, possibly including SOC. The change in total energy due to infinitesimal rotations of the magnetic moments is related to the Heisenberg exchange parameter $J_{ij}$ \cite{Liechtenstein1987} and DMI vectors $\vcc{D}_{ij}$ \cite{Udvardi2003,Ebert:DMIinKKR}. This approach is considered to give very accurate model parameters for large non-collinear magnetic textures.}
\item{(FLAPW--gBT) Secondly, we work in reciprocal space by means of spin-spiral states, Eq.~\eqref{eq:def_spinspiral}, employing the FLEUR code \cite{FLEUR} using the generalized Bloch theorem \cite{PhysRevB.69.024415} (gBT). Self-consistent calculations (without SOC) are performed for any arbitrary spin-spiral wave-vector $\vcc{q}$. We determine the $J_{ij}$ parameters from a fit to the spin-spiral energies for various spin-spiral vectors $\vcc{q}$ in the whole Brillouin zone (or high-symmetry lines thereof). The perturbative inclusion of SOC can be analogously related to the atomistic $\vcc{D}_{ij}$ parameters. This method is flexible in the sense that it allows to access both regimes of slowly and fast rotating non-collinear magnetic structures more realistically through self-consistent calculations.}
\item{(FLAPW--supercell) The third approach is conceptually similar to the FLAPW-gBT approach as it compares the energy of different spiraling magnetic states, but now in a supercell geometry. The direction of magnetic moments is fixed by constraining fields. In principle, no perturbative treatment of either SOC or the non-collinearity is needed, but due to the fast increase of computational costs with system size, this approach is practically restricted to small supercell sizes and hence large spin-spiral wave vectors $\vcc{q}$. We use two different magnetic configurations in a $(4\times1)$ supercell, one reproduces the setup of Yang \etal \cite{Yang2015a}, and the other employs small cone-angles to keep a nearly ferromagnetic alignment of Co-moments.}
\end{enumerate}

The parameters $A$ and $D$ for a micromagnetic model are either evaluated using Eqs.~\eqref{eq:stiffness:Jij} and \eqref{eq:spiralization:Dij} in case of the KKR method, or can be directly obtained from ab-initio calculations as quadratic and linear terms of the spin-spiral dispersion curve around the ferromagnetic state from FLAPW-gBT calculations \cite{Bode2007,Heide2008,Heide20092678,Zimmermann2014}.

In all approaches, we used the LDA exchange-correlation functional in the parameterization of Vosko \etal \cite{doi-10.1139/p80-159}. For the Co/Pt(111) system, we modeled the Pt substrate by 5 layers taking the structural relaxations as described in Sec.~\ref{section:structuralRelaxations}.

\subsubsection{The KKR method employing infinitesimal rotations}

We use the full-potential Korringa-Kohn-Rostoker Green function (FP-KKR-GF) method~\cite{juKKR,Bauer.phd} to converge the charge and spin densities in scalar-relativistic approximation for the ferromagnetic state using $30\times30$ $k$-points in the full Brillouin zone. In a next step, we obtain Heisenberg parameters $J_{ij}$ and DM vectors $\vcc{D}_{ij}$ in real-space by relating the change in energy of infinitesimal rotations of the magnetic moments at lattice sites $i$ and $j$ \cite{Liechtenstein1987,Ebert:DMIinKKR}, including spin-orbit coupling only in this step (one-shot SOC). Due to the infinitesimal rotations, this approach should be best to obtain parameters for large non-collinear structures, in particular skyrmions. We obtain interactions for Co pairs up to a distance of seven in-plane lattice constants and sample the full Brillouin zone by $160 \times 160$ $k$-points for this step. We truncate the expansion of the scattering wave-functions into spherical harmonics at $\ell_\mathrm{max}=3$. The energy contour integration includes a Fermi-function with an electronic temperature of 473~K, five Matsubara poles and is sampled by 39 points.

This methods also yields interaction parameters between strong Co moments and induced Pt moments. We can simply include these contributions in Eqs.~\eqref{eq:E_ss_from_Jij}, \eqref{eq:E_DM_from_Dij}, \eqref{eq:stiffness:Jij} and \eqref{eq:spiralization:Dij} which implicitly assumes that the size of the induced moments is rigid. A more sophisticated treatment is described in Appendix~\ref{appendix:inducedmoments}. As it turns out, the simple approach is sufficient for our considerations.

Another advantage of this method is the possibility to include SOC self-consistently in all steps and obtain an even more realistic set of parameters. However, the modification of the DM vectors is small (less than 0.1~meV for the nearest-neighbor interaction, $D_1$) and therefore a perturbative treatment of SOC is justified to obtain the DMI. We refer to Appendix~\ref{appendix:inducedmoments} for details.

\subsubsection{The FLAPW method employing spin-spiral states}

The second approach employs the FLAPW method as implemented in the DFT code FLEUR \cite{FLEUR} utilizing the generalized Bloch theorem (gBT)~\cite{PhysRevB.69.024415}. The main goal is to obtain the total energy of spin-spiral states, Eq.~\eqref{eq:def_spinspiral}, and extract parameters for a magnetic models thereof.
By virtue of the gBT, even long wave-length spin spirals meaning small values for $\abs{\vcc{q}}$ (the spin-spiral period length is given by $\lambda = 2\pi / \abs{\vcc{q}}$), can be treated very efficiently in the chemical unit cell, \ie without the need of large supercells.

Depending on the fitting procedure, we can either extract parameters for long wave-length magnon excitations (low-$\vcc{q}$ region) or for spin spiral ground states (large-$\vcc{q}$ region). To extract the Heisenberg $J_{ij}$ parameters, we converge the total energy of spin-spiral states in scalar-relativistic approximation using a $44 \times 44$ $k$-point mesh to an accuracy of 0.01~meV. The size of the basis set was determined by $K_{\mathrm{max}}=4.3~{a_\mathrm{B}}^{-1}$, with the Bohr radius $a_\mathrm{B}$.

In a second step, we include spin-orbit coupling (SOC) via first order perturbation theory~\cite{Heide20092678}. The energy correction $\Delta E_\mathrm{SOC}$ is used to extract either the micromagnetic DMI spiralization \cite{Freimuth2014} or the two-site DM vectors $\vcc{D}_{ij}$ for a spin-lattice model.

It is important to notice that the dispersion curves scale with the cone angle as by $\sin^2 \theta$. Hence, the energies for coned spin spirals as we use them later ($\theta=\pi/20$) are a factor of approx.\ 40 smaller as compared to flat spin-spirals, which requires quite high computational cutoffs.

\subsubsection{The supercell method}

\begin{figure}[thp]
\centering
\includegraphics[width=0.46\textwidth]{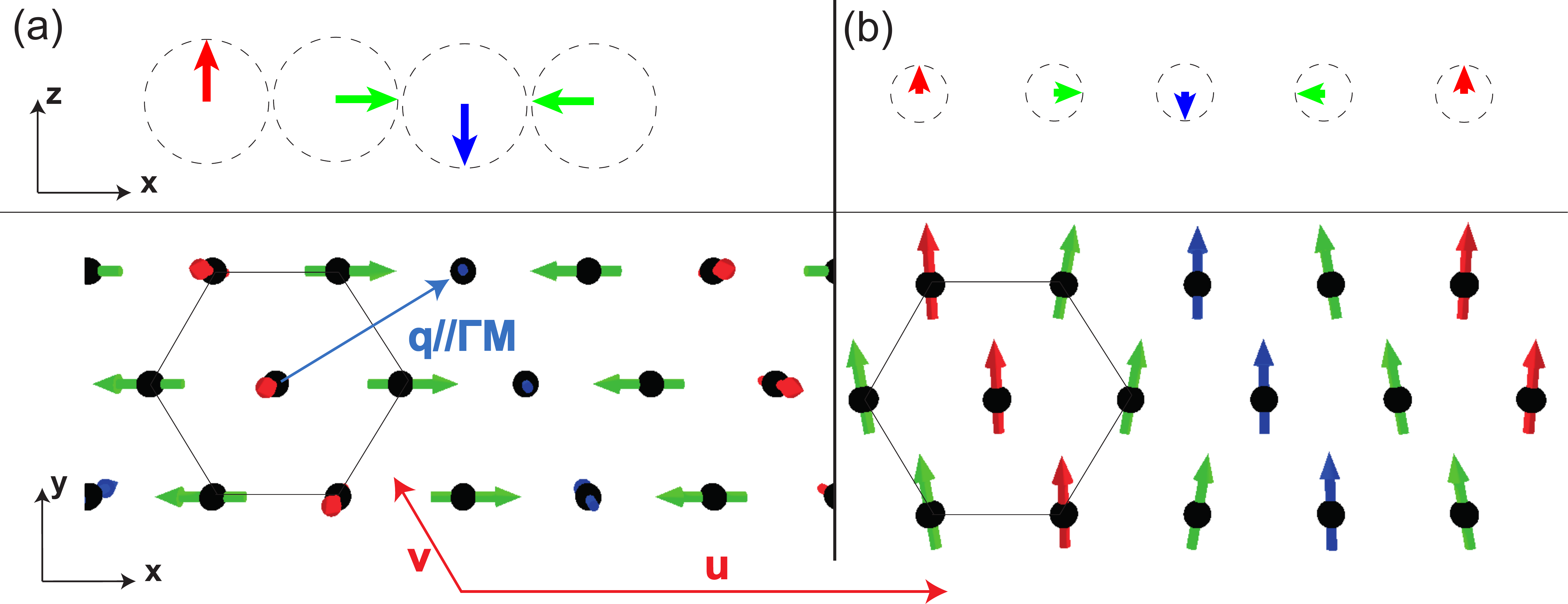}
\caption{Side and top-view onto two clockwise-rotating non-collinear spin configurations in a four-atom supercell (basis vectors are shown in red), representing (a) a flat spin-spiral where the magnetic moments are fully contained in the $xz$ plane and (b) a coned spin-spiral with small cone angle, where only a small component of the magnetic moments is contained in the $xz$ plane. In the latter case, the magnetic moments are almost pointing along the $y$ direction. The direction of the spin-spiral vector $\vcc{q}$ is indicated by a blue arrow.}
\label{fig:supercell}
\end{figure}

Additionally, we perform supercell calculations using the FLEUR code \cite{FLEUR} to extract the DMI. The supercell contains four magnetic atoms and is spanned by the basis vectors $\vcc{u}$ and $\vcc{v}$ (see Fig.~\ref{fig:supercell}).
We then impose magnetic states onto the supercell:
\begin{enumerate}
    \item{We mimic a flat spin-spiral (see Eq.~\eqref{eq:def_spinspiral} with $\theta=\pi/2$), \ie the direction of magnetic moments is given by $\vcc{m}_{i} = (\sin(i \pi/2), 0, \cos(i \pi/2))^\mathrm{T}$, where $i\in\{0,1,2,3\}$ labels the position of the magnetic atom in the supercell. This choice reproduces the setup of Yang~\etal~\cite{Yang2015a} and is shown in Fig.~\ref{fig:supercell}(a).}
    
    \item{We extend these states to coned spin-spirals [see Fig.~\ref{fig:supercell}(b)], where the direction of moments is given by
    \begin{eqnarray}
        \vcc{m}_i = \left( \begin{array}{c}
        \sin( \theta ) \sin\left(i \frac{\pi}{2} \right)  \\
        \cos( \theta )  \\
        \sin( \theta ) \cos\left(i \frac{\pi}{2} \right) \end{array}  \right), \quad i\in\{0,1,2,3\}
        \label{eq:m:coned_supercell}
    \end{eqnarray}
    where $\theta$ is the so-called cone-angle. The flat spin-spiral is reproduced for $\theta=\pi/2$, and all moments point along the $y$ direction for $\theta=0$.}
\end{enumerate}
In both cases, the lines of ferromagnetically aligned moments are parallel to the $\vcc{v}$ direction. This corresponds to spin-spiral states described by $\vcc{q} = \overline{\Gamma \mathrm{M}}/2$ [see Fig.~\ref{fig:supercell}(a)].

A nearest neighbor DMI, $D_\mathrm{eff}$, is derived by comparing the energies of right rotating [defined by Eq.~\eqref{eq:m:coned_supercell}] and left rotating magnetic structure [which is obtained by replacing $i \rightarrow -i$ in Eq.~\eqref{eq:m:coned_supercell}],
\begin{eqnarray}
\Delta E_{\mathrm{DMI}} = E_{\mathrm{right-SS}}-E_{\mathrm{left-SS}} = 24\, D_\mathrm{eff} \, \sin^2 \theta
\label{D_1:Yang}
\end{eqnarray}
We note that the parameter $D_\mathrm{eff}$ implicitly depends on the choice of $\theta$ through the self-consistent charge and spin densities which enter the Kohn-Sham Hamiltonian of DFT. On the one hand, we expect little modifications of the densities as compared to the FM state if $\theta \ll 1$. This configuration is very similar to a perturbative calculation in the vicinity of the FM state. On the other hand, we expect larger changes for $\theta = \pi/2$.

Concerning the computational details, a basis-set cutoff of $K_{\mathrm{max}}=4.1~a_\mathrm{B}^{-1}$ was chosen, and the Brillouin-zone was sampled by $12\times 48$ and $3\times12$ $k$-points for coned and flat spin-spirals, respectively. The spin-orbit interaction was included self-consistently in these calculations.

\section{Results}
\label{sec:results}

\subsection{Spin stiffness and Heisenberg exchange}
\label{sec:results:stiffness}

\begin{figure*}[thp]
\centering
\includegraphics[width=0.8\textwidth]{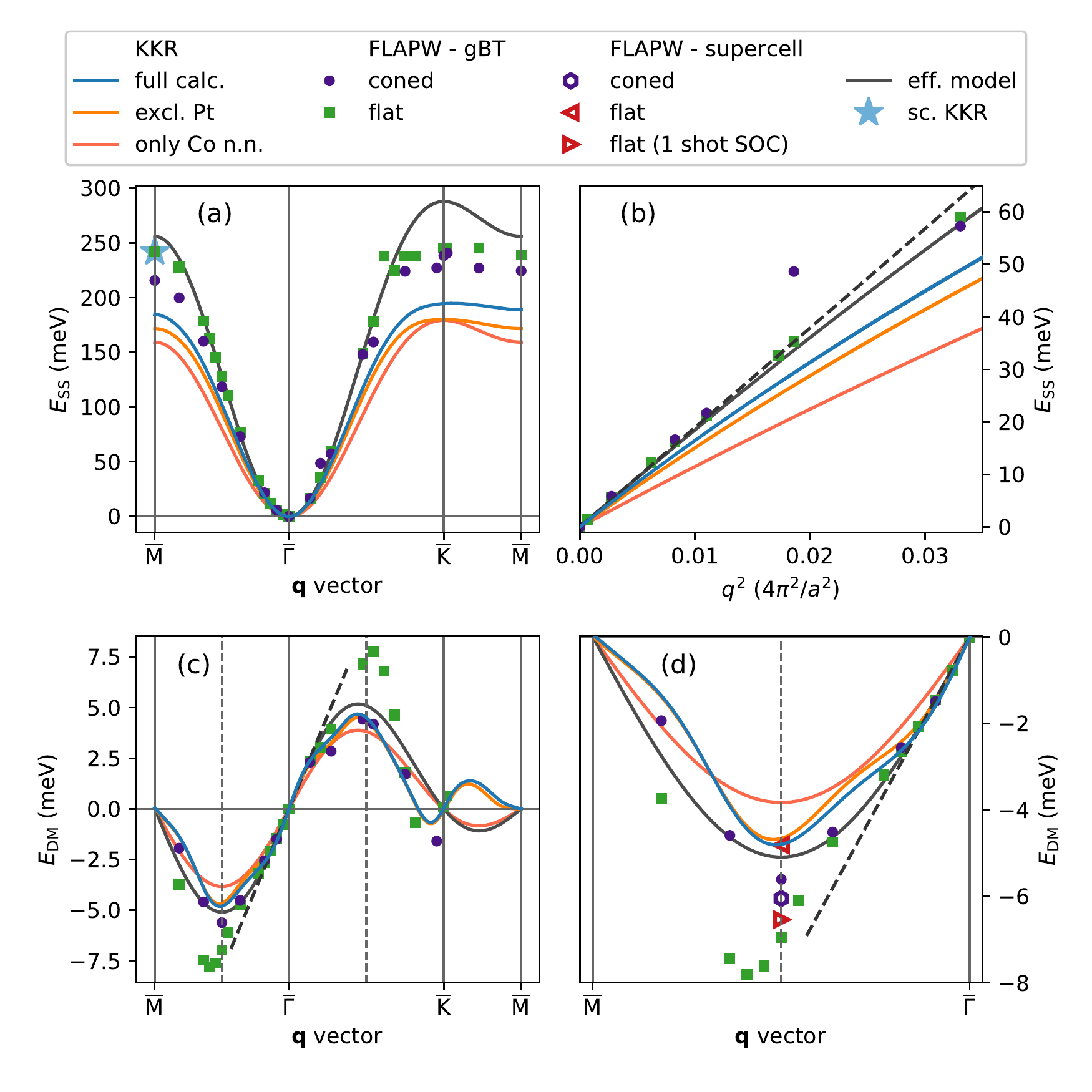}
\caption{Energy dispersion of spin-spirals for a monolayer Co(fcc) on Pt(111). (a) Non-relativistic dispersion curves $E_\mathrm{SS}(\vcc{q})$ along the high symmetry path of the first Brillouin zone. (b) Zoom into the parabolic region of (a) around the $\overline{\Gamma}$ point to obtain the spin stiffness. (c) Spin-orbit induced antisymmetric corrections $E_\mathrm{DM}(\vcc{q})$ to the energies along the high-symmetry path and (d) zoom onto the $\overline{\Gamma \mathrm{M}}$ direction. Full lines represent KKR-derived spin-spiral energies including all relevant interactions constants $J_{ij}$ (full calc.), exclude interactions between Co and Pt (excl. Pt) or consider only the nearest-neighbor Co interactions (only Co n.n.). Dashed lines indicate the slopes in the limit $\vcc{q}\rightarrow0$ and represent the (b) spin stiffness and (c-d) DMI spiralization. gBT = generalized Bloch theorem, 1 shot SOC = SOC included in last iteration only. For better comparison, the energies of coned spin-spirals have been scaled by $1/\sin^2 \theta$, where $\theta$ is the cone angle.}
\label{fig:dispersion-CoPt}
\end{figure*}

\begin{table*}[htb]
\centering
\caption{Coefficients $J_i$ and $D_1$ of the extended Heisenberg model, Eq.~ \eqref{eq:model_spinlattice}, the micromagnetic exchange stiffness $A$ and DMI spiralization $D$ evaluated for a Co-thickness of 0.2~nm, and the parameters of the effective nearest-neighbor model, $J_\mathrm{eff}$ and $D_\mathrm{eff}$. The FLAPW-gBT results are obtained from fits to the energy dispersion of coned and flat spin-spirals. The value for $D_1$ from FLAPW-gBT (flat) has been published in Ref.~\onlinecite{Dupe2014}. Both stacking positions of Co on the Pt(111) substrate (fcc or hcp) are considered. KKR-values in parentheses are obtained by an alternative evaluation method, which employs fitting of spin-spiral energies (see text and Ref.~\onlinecite{Simon2018}). For comparison, KKR results of Simon \etal \cite{Simon2018} (see also Ref.~\onlinecite{Vida2016}) are included, where a factor $1/2$ in $J_\mathrm{eff}$ and $D_\mathrm{eff}$ has been included due to a different definition of the spin-lattice Hamiltonian.}
\label{tab:coefficients}
\begin{ruledtabular}
\begin{tabular*}{\hsize}{cc@{\extracolsep{0ptplus1fil}}cccccccc}
{} & stacking & $J_1$ & $J_2$ & $J_3$ & $D_1$ & $A$ & $D$ & $J_\mathrm{eff}$ & $D_\mathrm{eff}$ \\
{} & position & (meV) & (meV) & (meV) & (meV) & (pJ/m) & (mJ/m$^2$) & (meV) & (meV) \\ \hline
KKR               & Co(fcc) & 19.9 & 1.7 & 0.4 & 1.1 & 40.9 (38.8) & 17.6 (14.76) & 29.5 (27.9) & 1.75 (1.47) \\
{}                & Co(hcp) & 20.8 & 1.5 &  0.2 & 1.0 & & & & \\[1.1ex]
FLAPW-gBT (coned) & Co(fcc) & 26.0 & 1.2 &  0.7 & 1.2 & 44.0 & 14.4 & 31.7 & 1.43 \\ 
{}                & Co(hcp) & 27.4 & 0.6 &  0.4 & 1.0 & & & & \\[1.1ex]
FLAPW-gBT (flat)  & Co(fcc) & 27.8 & 2.5 & -0.2 & 1.8 & 44.4 & 14.8 & 32.0 & 1.47 \\[1.1ex]
Ref.~\onlinecite{Simon2018} & Co(fcc) & & & & & (39.86) & (15.11) & (27.2) & (1.43) \\[1.1ex]
\end{tabular*}
\end{ruledtabular}
\end{table*}

Our main results are displayed in Fig.~\ref{fig:dispersion-CoPt} and summarized in Table~\ref{tab:coefficients}. Let us first compare the non-relativistic spin-spiral dispersion curves in Fig.~\ref{fig:dispersion-CoPt}(a-b): KKR and FLAPW data agree very well in the parabolic region around the $\overline{\Gamma}$ point, \ie in the region of interest when large non-collinear magnetic textures are studied. We zoom into this regime and plot the data as a function of $q^2$ in Fig.~\ref{fig:dispersion-CoPt}(b) to obtain the spin stiffness $A$ (or effective Heisenberg exchange $J_{\mathrm{eff}}$). Fits to the data of flat or coned spin-spirals yield the same spin stiffness of about 44~pJ/m, which converts to $J_\mathrm{eff} = 32$~meV (see Table~\ref{tab:coefficients}). The KKR-derived values are 10\% smaller as compared to FLAPW results. To estimate the role of the induced Pt moments for the energies from KKR, we switch their contributions manually off and see basically no change in the spin stiffness as compared to the full calculation. However, taking only the nearest-neighbor interactions of Co into account underestimates the spin stiffness by about 30\% [see Fig.~\ref{fig:dispersion-CoPt}(b)].

\begin{figure}[thp]
\centering
\includegraphics[width=0.48\textwidth]{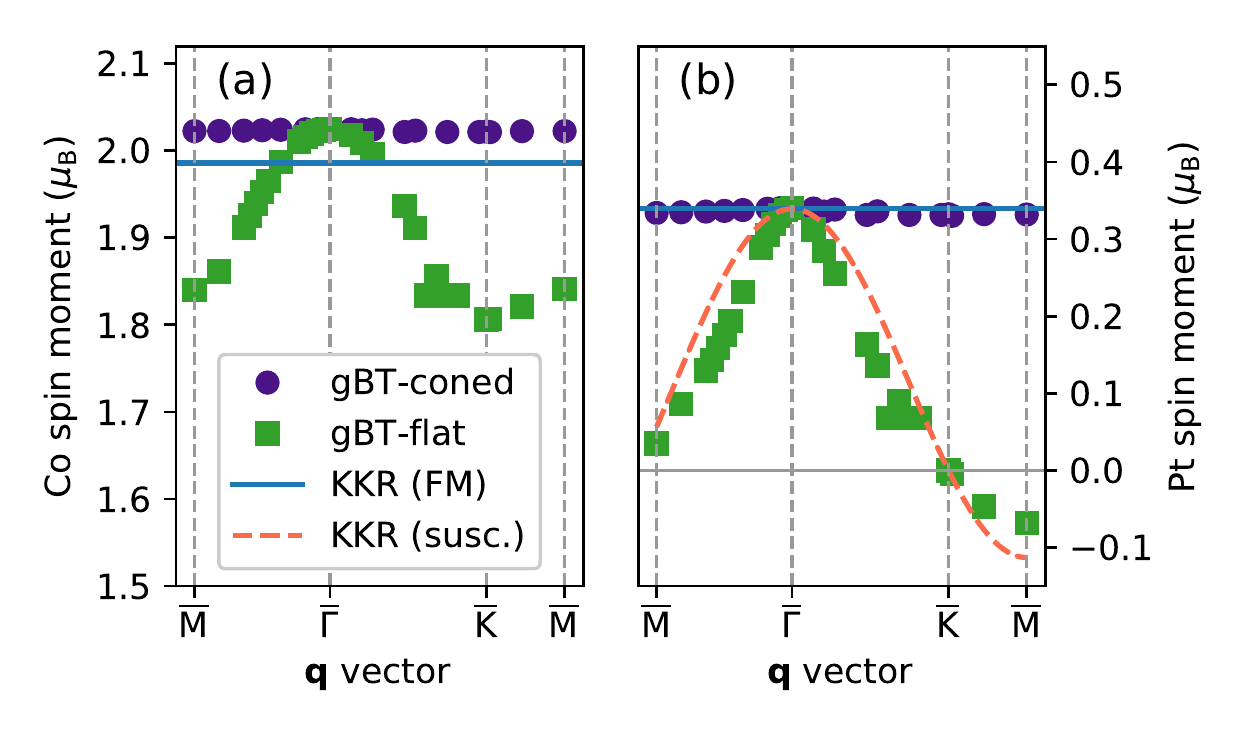}
\caption{Magnetic moments (a) for Co-atoms and (b) for induced Pt moments in the first substrate layer. The values from the ferromagnetic state of KKR calculations are shown as horizontal full lines. Induced Pt-moments as modeled by Eq.~\eqref{eq:Ptmoment:susceptibility} (see Appendix~\ref{appendix:inducedmoments}) are shown by the dashed line.}
\label{fig:magnetic_moments}
\end{figure}

As the $q$-vector approaches the Brillouin zone boundaries at the $\overline{\mathrm{M}}$ and $\overline{\mathrm{K}}$ points, larger deviations of about 70~meV in energy between flat spin-spirals from FLAPW-gBT calculations and KKR-deduced values occur [see Fig.~\ref{fig:dispersion-CoPt}(a)]. The reason lies partly in a different electronic structure: a frozen DFT effective potential from the ferromagnetic state (\ie $q$=0) is used in the KKR approach, whereas the potential for the spin-spiral is subject to change during the self-consistency as witnessed by a reduction of the Co moment (up to $-10\%$) and quenching of induced Pt-moments for large $q$ (see Fig.~\ref{fig:magnetic_moments}). In order to perturb the ferromagnetic state less, FLAPW-gBT calculations with coned spin-spirals are preferable. Indeed, the magnetic moments stay practically constant as function of spin-spiral vector $q$, and the spin-spiral energies get corrected to lower values towards the KKR result [see Fig.~\ref{fig:dispersion-CoPt}(a)], but a rather large discrepancy (of about \eg 30~meV at the $\overline{\mathrm{M}}$-point) remains. In order to test whether technical differences between the KKR and FLAPW methods, such as the division of space into Voronoi cells for KKR as opposed to muffin-tin spheres and interstitial region for FLAPW, induce such a discrepancy, we performed self-consistent KKR calculations for the $\overline{\mathrm{M}}$-point (i.e. the row-wise anti-ferromagnetic state) in a simple $(2 \times 1)$-supercell geometry. We obtain a perfect agreement to the FLAPW result [compare the green square on top of the blue star at the $\overline{\mathrm{M}}$ point in Fig.~\ref{fig:dispersion-CoPt}(a)], highlighting the consistency of the two methods if similar approximations are employed. We can finally speculate that the energy difference of about 30~meV at the $\overline{\mathrm{M}}$-point between coned spin-spiral and KKR-deduced energies might be caused by higher order magnetic exchange interactions, such as the four-spin-three-site-interaction\cite{Kronlein2018}, which are not captured by the KKR approach based on infinitesimal rotations, but naturally included in the self-consistent spin-spiral calculations. If present, it leads to an effective renormalization of the $J_{ij}$'s (concretely, $J_1$ would be increased by 30\%), which also reflects in an increased Curie temperature as compared to KKR (see Appendix~\ref{appendix:MonteCarlo}).

\subsection{Dzyaloshinskii-Moriya interaction}

Let us now turn to the DMI [see Figs.~\ref{fig:dispersion-CoPt}(c-d)]. We obtain a very good agreement throughout the whole Brillouin zone between spin-spiral energies from KKR and FLAPW-gBT calculations with coned spirals (compare blue line and purple dots). Flat spin-spirals yield considerably higher DM-energies for $q$-vectors in the middle of $\overline{\Gamma \mathrm{K}}$ and $\overline{\Gamma \mathrm{M}}$, respectively. This discrepancy is again a reflection of the different electronic structure for large $q$-vectors and links directly to the changed hybridization between Co and Pt \cite{Sandratskii2017}. We can conclude that it might fail to take a single-point calculation from a rather large $q$ value and infer information about the low-$q$ regime from this point. In the micromagnetic limit, however, coned and flat FLAPW-gBT spin-spiral calculations yield a similar DMI of $D = 14.4~\mathrm{mJ/m^2}$ (which converts to $D_\mathrm{eff}=1.43~\mathrm{meV}$; see dashed line). The KKR-obtained $E_\mathrm{DM}$ data is also well approximated by this line [see Fig.~\ref{fig:dispersion-CoPt}(d)]. If we carefully converge the DMI spiralization from KKR according to Eq.~\eqref{eq:spiralization:Dij} with respect to the number of neighbors, we obtain a 20\% (10\%) higher DMI as compared to FLAPW-gBT values including (neglecting) the interactions between Co and induced Pt moments. This enhancement is due to the fact that by evaluating Eq.~\eqref{eq:spiralization:Dij}, we calculate the exact ($q \rightarrow 0$)-limit. Indeed, if we perform a linear fit to the KKR-derived DMI energies in an interval of $\abs{q} \leq 0.1~\frac{2\pi}{a}$, similarly to the procedure used for FLAPW-gBT calculations, we obtain an identical spiralization (up to two digits) as compared to the FLAPW-gBT calculations. Simon \etal \cite{Simon2018} use an analogous fitting-procedure with pair-interaction parameters as calculated by the KKR method, employing a different code, and obtain a value in perfect agreement to our KKR-derived values (see Table~\ref{tab:coefficients}).

The fact that our DMI data exhibits the same slope from $\overline{\Gamma}$ towards $\overline{\mathrm{M}}$ and $\overline{\mathrm{K}}$, respectively, shows that the DMI spiralization is isotropic, which is expected for a system with $C_{3v}$ symmetry \cite{Bogdanov89}. The positive sign of the slope corresponds to a lowering of energies of magnetic states with left-handed (anti-clockwise) chirality.

The effective model, which is obtained from the fits in the low-$q$ region, reproduces the energies from coned spin-spirals and via KKR rather satisfactorily even for large $q$ (see solid black line in Fig.~\ref{fig:dispersion-CoPt}c). We observe pronounced deviations from the simple sine-behavior in the middle of the Brillouin zone for the KKR-obtained $E_\mathrm{DM}$ curves, which stem from contributions beyond the nearest neighbors.
Near the $\overline{\mathrm{K}}$-point (which represents the non-collinear $120^\circ$ N\'{e}el state for a flat spin-spiral), a qualitative change of the energy dispersion is caused by interactions beyond nearest neighbors and cannot be captured by the effective model. Similarly, a different slope between effective model and KKR-derived energy curves appears near the $\overline{\mathrm{M}}$-point, highlighting the limitations of the effective model when extrapolating from the low-$q$ region to arbitrary $q$-vectors.

\begin{table}[htb]
\centering
\caption{DMI values as extracted from FLAPW supercell calculations using a coned or flat spin spiral, either treating SOC self-consistently (scSOC) or including it in the last iteration only (1-shot SOC). The Co-monolayer is in the fcc-stacking position on the Pt(111) substrate. For comparison, results of Yang \etal \cite{Yang2015a} (flat spirals, scSOC, using the plane-wave code VASP) are included, where a factor $1/2$ in $D_\mathrm{eff}$ has been included due to a different definition of the spin-lattice Hamiltonian.}
\label{tab:supercell_DMI}
\begin{ruledtabular}
\begin{tabular*}{\hsize}{cc@{\extracolsep{0ptplus1fil}}cc}
{}    & {}         & $D$        & $D_\mathrm{eff}$ \\
{}    & {}         & (mJ/m$^2$) & (meV) \\ \hline
coned & scSOC      & 20.3       &  2.02 \\[1.1ex]
flat  & scSOC      & 16.2       &  1.60 \\
{}    & 1-shot SOC & 21.9       &  2.18 \\[1.1ex]
Ref.~\onlinecite{Yang2015a} & & 19.0 & 2.17 \\
\end{tabular*}
\end{ruledtabular}
\end{table}

Next, we discuss the Dzyaloshinskii-Moriya interaction as determined from supercell calculations and take the effective model with values as fitted from FLAPW-gBT calculations as reference. A coned supercell-spiral yields a DM energy very close to the corresponding FLAPW-gBT calculation at $\overline{\Gamma \mathrm{M}}/2$ [see Fig.~\ref{fig:dispersion-CoPt}(d)]. Despite this good agreement for this single point (5.7 vs.\ 6.1~meV), the inferred $D_\mathrm{eff}$ is 40\% higher as compared to the values from fits in the low-$q$ region of FLAPW-gBT calculations (see Table~\ref{tab:supercell_DMI}), which again emphasizes the difficulty of relying on only one data point. The DMI coefficient of a flat supercell-spiral agrees surprisingly well with our reference value (1.60 as compared to 1.43~meV). However, the corresponding energy $E_\mathrm{DM}$ is considerably lower than the spin-spiral dispersion of FLAPW-gBT calculations for this state [see Fig.~\ref{fig:dispersion-CoPt}(d)]. We can trace this discrepancy back to the different treatment of SOC, which is included self-consistently in the supercell calculations but only in first-order perturbation theory in the FLAPW-gBT calculations: if we include SOC in the supercell calculation by a force-theorem approach, \ie performing a single iteration with SOC on top of a converged scalar-relativistic calculation and compare the sum of single-particle energies, the DM energy agrees within 5\% with the gBT calculation (6.5 vs.\ 6.8~meV). Similar findings have been reported for freestanding Fe/Ir bilayers \cite{Meyer2017}. The energies of Yang \etal \cite{Yang2015a} (taking their values from Co(1)/Pt(3), where the number in parentheses denotes the number of atomic layers) are 35\% higher than our corresponding supercell calculation (compare 2.17 to 1.60~meV), probably due to the different number of substrate layers and different relaxations of atomic positions, and about 50\% higher as our reference value from the low-$q$ regime. The good agreement between the value of Ref.~\onlinecite{Yang2015a} and our 1-shot SOC supercell calculation (see Table~\ref{tab:supercell_DMI}) is due to an accidental cancellation of errors. Note that the authors of Ref.~\onlinecite{Yang2015a} used a different magnetic volume (corresponding to the volume an atom in fcc-Pt) for the conversion between $D$ and $D_\mathrm{eff}$ (\cf Eq.~\eqref{eq:conversion:Deff+D}), which leads to a different ratio of $D_\mathrm{eff}/D$ as compared to us.

\subsection{Discussion}

Overall, our results emphasize the compatibility of the FLAPW-gBT and KKR methods in the relevant micromagnetic limit, as well as the flexibility of the generalized Bloch theorem approach as it can access states beyond the micromagnetic limit more realistically through self-consistent calculations. Our findings regarding the DMI for the present case of a Co monolayer on Pt(111) are in satisfactory agreement to previous studies on this system \cite{Simon2018,Yang2015a,Zimmermann2018}.

Comparing our value for the spin stiffness to the literature, it is considerably higher than most of the experimentally determined values for Co: for bulk-Co a stiffness of about 15~pJ/m (fcc) and 30~pJ/m (hcp-Co) is measured by various methods \cite{Huller1986}, a value of 21~pJ/m was measured for 10~nm thick Co-films \cite{Eyrich2012}, and even smaller stiffnesses as low as 14~pJ/m are reported for ultra-thin Co films down to thicknesses of 0.5~nm. The reported spin stiffness by Boulle \etal \cite{Boulle2016} represents an exception to this trend, as they determine 27.5~pJ/m for Co-thicknesses below 1~nm. Still, our results are nearly 50\% higher than this value, and they are in line with previous DFT calculations on a Co monolayer on Pt(111)~\cite{Vida2016,Simon2018}.

One possibility for the discrepancy between experiment and theory might arise from the fact that experiments are necessarily performed at finite temperatures, whereas the ab-initio calculations neglect any spin-fluctuations. In simplest approximation, the spin-stiffness $A$ and DMI spiralization $D$ are renormalized by $(M(T)/M_0)^2$, where $M(T)$ is the temperature-dependent magnetization and $M_0$ the saturation magnetization \cite{Rosza2017}. Inserting the values for $M(T)/M_0$ at $T=300$~K as deduced from Monte Carlo simulations [see Appendix~\ref{appendix:MonteCarlo} and Fig.~\ref{fig:thermo}(b)], we arrive at a renormalization factor of 0.67 and a room-temperature spin-stiffness of 29.7~pJ/m (using results based on parameters from FLAPW-gBT calculations), which is in much better agreement with the experimentally determined spin-stiffnesses.

The critical temperature at which the magnetization vanishes, $T_\mathrm{c}$, was also extracted from the Monte Carlo simulations. Depending on the parameterization, these values are ranging from 500~K to 620~K for the KKR and the FLAPW-gBT parameterization, respectively.
These critical temperatures seem to be quite high for a two-dimensional (2D) system: They are up to 40\% higher than one would estimate based on the Curie temperature of bulk Co, $T_\mathrm{c}^\mathrm{(3D)}=1400$~K, as calculated by
\begin{equation}
  T_\mathrm{c}^\mathrm{(2D, est.)} = \frac{2~T_\mathrm{c}^\mathrm{(3D)}}{\ln\left( \frac{3\pi}{4} \frac{k_\mathrm{B}~T_\mathrm{c}^\mathrm{(3D)}}{K} \right)} \approx 440 ~\mathrm{K}, \label{eq:Tc:estimate}
\end{equation}
with Boltzmann's constant $k_\mathrm{B}$ and uniaxial anisotropy $K=0.5$~meV. Eq.~\eqref{eq:Tc:estimate} is based on an renormalization group analysis and assumes unchanged magnetic interaction parameters when going from 3D to 2D (see Ref.~\onlinecite{Bluegel_Magnetism_low_dimensions} and references therein for details). But we infer that $J_1=19.9$~meV of a monolayer Co/Pt(111) is about 50\% higher as compared to hcp-Co, which we determined to $J_1=13$~meV by the KKR method, in very good agreement to 14.8~meV as determined by LMTO calculations~\cite{Pajda2001}. This increased $J_1$ explains the discrepancy between Eq.~\eqref{eq:Tc:estimate} and Monte-Carlo simulations for the monolayer. This difference of the exchange interactions stems from changes in the electronic structure and corrects the estimated value to higher temperatures. Inserting in Eq.~\eqref{eq:Tc:estimate} the Co-bulk Curie temperature of a hypothetical Co solid with an increased Curie temperature of $\tilde{T}_\mathrm{c}^\mathrm{(3D)}=(19.9/13) ~ T_\mathrm{c}^\mathrm{(3D)}=2140$~K corresponding to the nearest neighbor exchange interaction of $J_1=19.9$~meV we obtain an estimated $\tilde{T}_c^\mathrm{(2D, est.)}$ of 630~K.

Indeed, this corrected estimate and the Monte-Carlo determined values are well compatible to magneto-optical Kerr effect experiments on a monolayer Co on Pt(111)\cite{Shern1999}: a finite magnetization is present up to 623~K. However, the experimental situation is more complex, as the formation of a CoPt surface-alloy is observed above 500~K, and the reported 623~K is an increased value compared to the ideal Co-monolayer case\cite{Shern1999}. Consequently, the critical temperature of the ideal Co monolayer on Pt(111) lies in between 500--623~K. All our Monte-Carlo-predicted critical temperatures are compatible with the experiment, whereas $T_\mathrm{c}^\mathrm{(2D, est.)}$ estimated from the bulk properties is not. This highlights that a prediction of the Curie temperature of thin films purely from the knowledge of bulk properties is to be taken with caution and possibly fails. Overall, our calculations confirm that the Curie temperature of Co thin-films on Pt(111) lies far above room temperature\cite{Chappert1998,Hashimoto1990}, even for the monolayer.

Having obtained consistent values for the spin-stiffness and DMI spiralization, and considering additionally the uniaxial anisotropy $K_i=0.5~\mathrm{meV}$ (which converts to $K=6~\mathrm{MJ}/\mathrm{m}^3$), which we obtained by FLAPW calculations with out-of-plane easy axis, we can determine the ground state. Due to the rather strong magnetocrystalline anisotropy, we obtain a ferromagnetic ground state\footnote{The quantity $\kappa = \frac{16}{\pi^2}\frac{AK}{D^2}$ denotes a ferromagnetic ground-state for $\kappa > 1$ \cite{Zimmermann2014}. With our values for Co/Pt(111), we obtain $\kappa \approx 2$.}. We can in hindsight check the validity of the micromagnetic limit by estimating the domain-wall width
\begin{equation}
  w = 2 \sqrt{\frac{A}{K}} = 5.4~\mathrm{nm}
\end{equation}
and with the arguments presented in Appendix~\ref{appendix:lengthscales}, we are indeed in the micromagnetic limit ( $\vcc{q} < 0.03~\frac{2\pi}{a}$). Our calculated domain-wall width is in perfect agreement with the experimental value of about 4~nm in monolayer Co/Pt(111) as measured by spin-polarized scanning-tunneling microscopy (SP-STM) at low temperatures \cite{Meier2006}. Very recently, a domain-wall width of $17.2$~nm in slightly thicker (1.4~nm) epitaxial Co layers on Pt(111) has been measured by SEMPA at room temperature \cite{Corredor2017}.

Another characteristic length scale for spiraling magnetic textures induced by DMI is set by $L_\mathrm{D} = 4\pi A / D \approx 40~\mathrm{nm}$, taking the values from Table~\ref{tab:coefficients}, which is even an order of magnitude larger than the domain-wall width and approximates the micromagnetic limit even more.

Finally, we have determined the magnetic interaction parameters for the spin-lattice model for the hcp stacking position of Co on Pt(111) and present them in Table~\ref{tab:coefficients}. The changes are marginal as compared to the fcc stacking position, in agreement to Ref.~\onlinecite{Vida2016}.

\section{Conclusion}
\label{sec:conclusions}

We have determined the magnetic interaction parameters for a Co monolayer on Pt(111) by three distinct approaches, (i) performing infinitesimal rotations, (ii) using spin-spirals employing the generalized Bloch theorem  for various $q$ vectors,  (iii) and constraining spin-spirals into a rather small supercell. We obtain consistent results for the spin stiffness and the Dzyaloshinskii-Moriya interaction in the long-wavelength (micromagnetic) limit around the ferromagnetic state using methods (i) and (ii). When going to higher spin-spiral $q$ vectors, deviations through differences in the electronic structure play a role for flat spin spirals. In order to still realize the long-wavelength limit in the supercell approach, we propose to use a coned spiraling structure with a small cone angle, which leaves \eg the magnetic moments unchanged as function of the spin-spiral $q$ vector (being inversely proportional to the supercell size). We found that the micromagnetic DMI spiralization might not be accurately inferred by extrapolating from one data point obtained for a large $q$-vector (\ie small supercell).

\section*{Acknowledgments}
We acknowledge discussions with Stanislas Rohart. B.\ Dup\'e thanks DARPA TEE program (\#HR0011727183) and the (DFG) via the project DU 1489/3-1 for financial support. B.\ Dup\'e and S.\ Heinze thank the Deutsche Forschungsgemeinschaft (DFG) for financial support via the project DU 1489/2-1.  B.\ Zimmermann, S.\ Heinze and S.\ Bl\"ugel thank the European Unions Horizon 2020 research and innovation programme under grant agreement No.\ 665095 (FET-Open project MAGicSky) and S.\ Bl\"ugel thanks DARPA TEE program (\#HR0011831554) from DOI, Deutsche  For\-schungs\-gemeinschaft (DFG) through SPP 2137 ``Skyrmionics", and the Collaborative Research Center SFB 1238 (Project C1) for funding. S.\ Lounis and M. Bouhassoune acknowledge support from the European Research Council (ERC) under the European Union's Horizon 2020 research and innovation program (ERC-consolidator grant 681405 -- DYNASORE). We gratefully acknowledge computing time at the HLRN supercomputer as well as JURECA supercomputer (projects jias1a and jias1f) of J\"ulich Supercomputing Centre and JARA-HPC of RWTH Aachen University.
\appendix

\section{Thermodynamical study of Co/Pt(111)}
\label{appendix:MonteCarlo}

\begin{figure}[thp]
\centering
\includegraphics[width=0.48\textwidth]{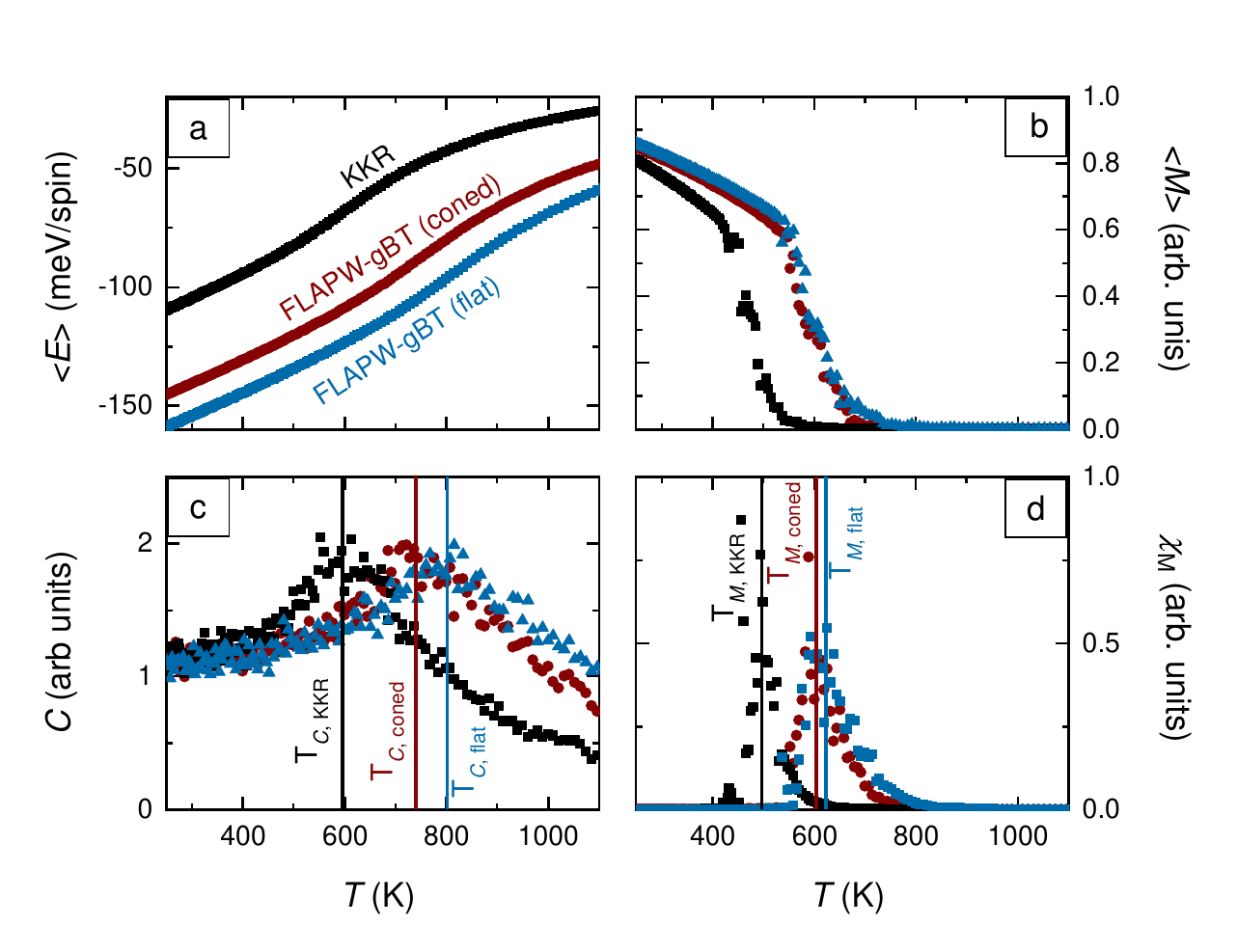}
\caption{Calculation of the critical temperature of Co/Pt(111) for the different parameterization of the extended Heisenberg Hamiltonian from Monte Carlo simulations. (a) Averaged total energy, (b) magnetization, (c) specific heat and (d) magnetization susceptibility as a function of temperature.}
\label{fig:thermo}
\end{figure}
 
As discussed in the main text (see Sec.~\ref{sec:results:stiffness}), the KKR as well as FLAPW-gBT methods employing flat or coned spin-spirals all leave the curvature of the spin-spiral dispersion around the $\overline{\Gamma}$-point unchanged. However, large changes occur for larger $q$-vectors corresponding to strongly non-collinear magnetic textures. These states will only be present at high temperature or after an exciting stimulus with high energy. Therefore, to understand the consequences of the different  parameterizations of the Heisenberg Hamiltonian on these states, we have performed atomistic Monte Carlo simulations and calculated the critical temperature T$_{\mathrm{C}}$ of a Co monolayer in the fcc stacking position on Pt(111).

We took the parameterization of the extended Heisenberg model as determined from KKR and FLAPW-gBT calculations (\cf Table~\ref{tab:coefficients}), and included an uniaxial anisotropy of $K_i=0.5$~meV which was determined by FLAPW calculations. Subsequent Monte Carlo simulations have been performed using a $300 \times 300$ supercell on 192 replicas of the supercell with temperatures geometrically spaced between $200$ and $1400$ Kelvin. During the Metropolis steps, we have averaged over 600 steps the total energy $\langle E \rangle$ as obtained from Eq.~\eqref{Eq:H}, the total magnetization $\langle M \rangle$ as well as the specific heat $C$ and the magnetization susceptibility $\chi_M$. 

Fig.~\ref{fig:thermo} shows the results of the Monte Carlo simulations for the three different parameterizations. The black, red and blue curves correspond to the parameterization obtained with via KKR and FLAPW-gBT methods employing coned and flat spin spirals, respectively. Fig.~\ref{fig:thermo}(a) shows the averaged total energy as a function of temperature. All energies show a smooth increase with a change of slope at around 500~K which indicates a phase transition. As the total energy of the magnetic lattice increases, the magnetization decreases as displayed in Fig.~\ref{fig:thermo}(b): The magnetization decreases linearly up to the critical temperature and then drops abruptly down to zero. The drop occurs at a temperature of 600~K for both FLAPW-gBT parameterizations and at 500~K for the KKR parameterization.

To characterize the phase transition in more detail, we calculate the specific heat $C$ and the magnetic susceptibility $\chi_M$ [\cf Fig.~\ref{fig:thermo}(c) and (d), respectively] as explained in Ref.~[\onlinecite{Bottcher2017}]. When DMI is present in a magnetic system, the phase diagram shows an intermediate region between the paramagnetic phase at high temperature and the long-range ordered phase at low temperature. In this intermediate regime, the number of skyrmions can fluctuate \cite{Rozsa2016,Hou2017a} and may be used to nucleate skyrmions in magnetic multilayers by current pulses \cite{Lemesh2018}. This region is present between the critical temperature extracted from the location of the peak of the specific heat, $T_C$, and the critical temperature extracted from the peak of the magnetization susceptibility, $T_M$ \cite{Bottcher2017}.

The specific heat curves show a peak at 600~K for the parameterization from KKR and at 740~K and 800~K for the ones from FLAPW-gBT employing conical and flat spin spirals, respectively. The subsequent phase transition to the long-range ordered phase occurs at 500~K (KKR), 600~K (FLAPT-gBT coned) and 620 (FLAPW-gBT flat). The predicted transition temperatures from the KKR parameterization are lower than the corresponding ones from FLAPW-gBT, with differences of up to 200~K (or about 25\%).

\section{Arguments for the micromagnetic limit for Co/Pt interfaces} \label{appendix:lengthscales}

In order to determine the relevant range of spin-spiral vectors $q$ for Co/Pt, we perform a power spectrum analysis of the well known domain wall profile. The magnetization of a N\'{e}el-type domain wall separating the ferromagnetic domains $\vcc{m}(x \rightarrow \pm \infty)= (0,0,\pm 1)^\mathrm{T}$ is given by $\vcc{m}(x) = (\cos \Theta, 0, \sin \Theta)^\mathrm{T}$, with
\begin{equation}
    \Theta(x) = \arcsin \tanh \left( \frac{x}{w/2} \right) ~,
\end{equation}
where $w$ is the domain-wall width.

\begin{figure}[thp]
\centering
\includegraphics[width=0.45\textwidth]{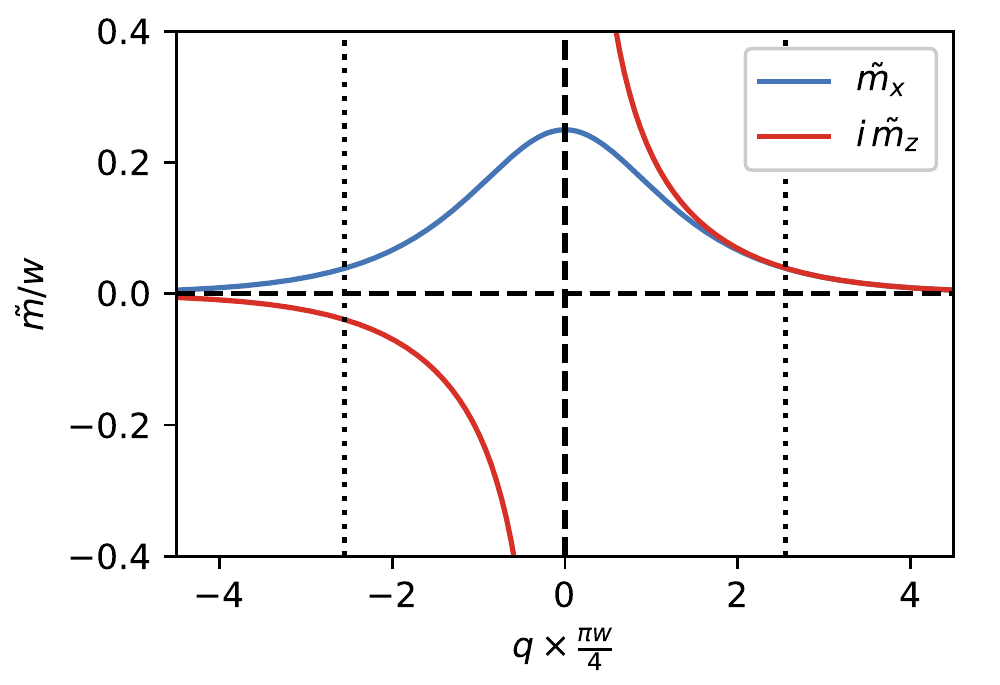}
\caption{Power spectrum of the magnetization profile for a N\'{e}el-type domain wall. Vertical dotted lines indicate the interval that contains 90\% of the components $\tilde{m}_x$.}
\label{fig:DW_power_profile}
\end{figure}

The Fourier transformation of the $m_x$ component is directly evaluated as
\begin{equation}
    \tilde{m}_x(q) = \frac{1}{2\pi} \int \mathrm{d}x ~ {m}_x(x) ~ \mathrm{e}^{-i q x} = \frac{w}{4 \cosh(\pi w q/4)} ~.
\end{equation}
For the $m_z$-component, we need to regularize the Fourier transformation and obtain
\begin{eqnarray}
    \tilde{m}_z(q) &=& \frac{1}{2\pi} \lim_{\epsilon \rightarrow 0^+}  \int \mathrm{d}x ~ {m}_z(x) ~ \mathrm{e}^{-i q x} ~ \mathrm{e}^{-\epsilon x} \\
    &=& -i \frac{w}{4 \sinh(\pi w q/4)} ~.
\end{eqnarray}
Both Fourier transformed components peak around $q=0$ (see Fig.~\ref{fig:DW_power_profile}), and are significant only for
\begin{equation}
    \left\lvert q \frac{\pi w}{4} \right\rvert \leq 2.55.
\end{equation}
This interval has been chosen such that it contains more than 90\% of the components $\tilde{m}_x$. Inserting a typical value for the domain-wall width in Co/Pt based systems ($w \approx 5~\mathrm{nm}$) yields
\begin{equation}
    \lvert q \rvert \leq 0.03 ~ \frac{2\pi}{a},
\end{equation}
with the in-plane lattice constant of the Co/Pt(111) film, $a=0.276~\mathrm{nm}$. These very low $q$ values justify the application of the micromagnetic approach for Co/Pt based systems.

\section{Treatment of induced Pt moments in non-collinear structures and self-consistent inclusion of SOC}
\label{appendix:inducedmoments}

The size of the induced Pt moment can vary strongly in non-collinear structures, as shown by results for flat spin-spirals in Fig.~\ref{fig:magnetic_moments}(b). To account for such a variation in the KKR calculations, a scheme based on the non-local spin-susceptibility was proposed \cite{Polesya2010}, where the size and direction of the induced Pt moment $\vcc{M}_\mathrm{Pt}$ is determined by the orientation of the nearest-neighbor Co-moments,
\begin{eqnarray}
    \vcc{M}_\mathrm{Pt} &=& \chi^\text{Co-Pt} \sum_{j\in \text{n.n}} \vcc{M}^j_\mathrm{Co} ~ , \label{eq:Ptmoment:susceptibility} \\
    \chi^\text{Co-Pt} &=& \frac{\abs{\vcc{M}^\mathrm{FM}_\mathrm{Pt}}}{\sum_{j\in \text{n.n}} \abs{\vcc{M}^j_\mathrm{Co}}}~.
\end{eqnarray}
The superscript `FM' denoted the moment obtained for the ferromagnetic state.

The induced Pt moment approximated in this way with a values for $\abs{\vcc{M}^\mathrm{FM}_\mathrm{Pt}}=0.34~\mbohr$ as calculated by KKR reproduces the results from flat spin-spiral calculations using the FLAPW-gBT method rather well [compare the dashed line and green squares in Fig.~\ref{fig:magnetic_moments}(b) in the main text]. Remaining deviations are probably due to the variation of the size of the Co moments in FLAPW-gBT calculations (see Fig.~\ref{fig:magnetic_moments}(a)].

\begin{figure}[thp]
\centering
\includegraphics[width=0.45\textwidth]{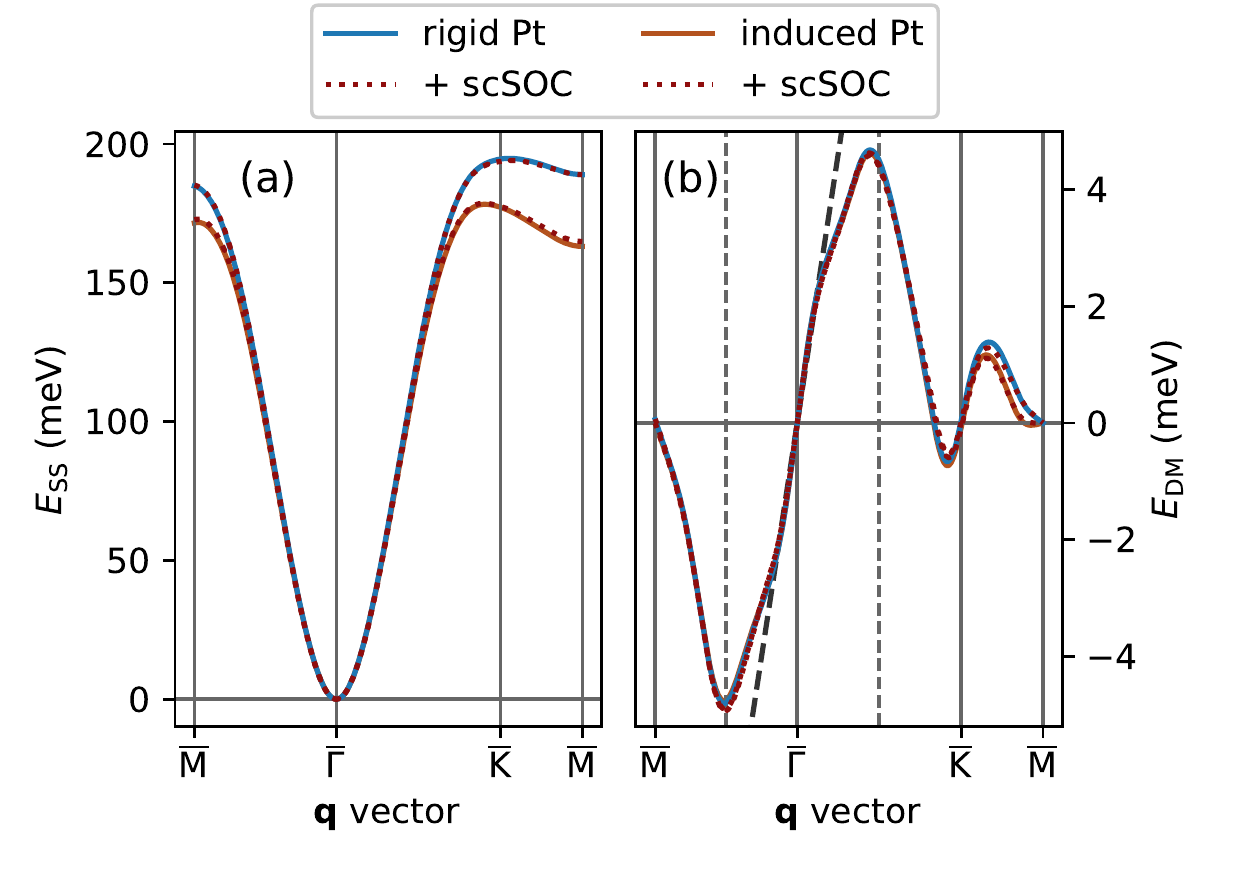}
\caption{Energy dispersions for spin-spirals as evaluated with parameters as obtained by the KKR method for (a) the exchange interaction and (b) Dzyaloshinskii-Moriya interaction. The contributions from interactions between Co and induced Pt moments are included assuming either rigid Pt moments or a variation according to Eq.~\eqref{eq:Ptmoment:susceptibility}.}
\label{fig:Energies:test:inducedPt}
\end{figure}

As a next step, we can evaluate the spin-spiral dispersion curves by renormalizing the KKR-obtained parameters $J^\text{Co-Pt}$ and $\vcc{D}^\text{Co-Pt}$ by $\abs{\vcc{M}_\mathrm{Pt}}/\abs{\vcc{M}^\mathrm{FM}_\mathrm{Pt}}$. However, this treatment does not significantly change the energies, as shown in Fig.~\ref{fig:Energies:test:inducedPt}. In particular, the relevant low-$q$ regime around $\overline{\Gamma}$ is not affected.

We also tested whether the inclusion of SOC in the self-consistent determination of the charge and spin-density affects the $J_{ij}$ and $\vcc{D}_{ij}$ parameters as determined by infinitesimal rotations. The resulting changes in $A$ and $D$ (or $J_\mathrm{eff}$ and $D_\mathrm{eff}$ are 3\% at most, and derived spin-spiral energies are hardly distinguishable as compared to the (1-shot SOC)-procedure employed in the main text (compare dotted to full lines in Fig.~\ref{fig:Energies:test:inducedPt}).

\bibliography{refs}

\end{document}